\documentclass[prb,twocolumn,showpacs,aps,superscriptaddress,floatfix]{revtex4-2}
\usepackage{amsmath}
\usepackage{amssymb}
\usepackage{amsfonts}
\usepackage{bm}
\usepackage[dvips]{graphicx}
\usepackage{color}
\usepackage{ulem}
\usepackage[unicode, pdftex]{hyperref}
\usepackage{verbatim}
\begin{document}
\title{Born approximation study of the strong disorder in magnetized surface states of topological insulator}
\author{R.S. Akzyanov}
\affiliation{Dukhov Research Institute of Automatics, Moscow, 127055 Russia}
\affiliation{Moscow Institute of Physics and Technology, Dolgoprudny,
    Moscow Region, 141700 Russia}
\affiliation{Institute for Theoretical and Applied Electrodynamics, Russian
    Academy of Sciences, Moscow, 125412 Russia}

\begin{abstract}
In this study we investigate the effect of random point disorder on the surface states of a topological insulator with out-of-plane magnetization. We consider the disorder within a high order Born approximation. The Born series converges to the one branch of the self-consistent Born approximation (SCBA) solution at low disorder. As the disorder strength increases, the Born series converges to another SCBA solution with the finite density of states within the magnetization induced gap. Further increase of the disorder strength leads to a divergence of the Born series, showing the limits of the applicability of the Born approximation. We find that the convergence properties of this Born series are closely related to the properties of the logistic map, which is known as a prototypical model of chaos. 
We also calculate the longitudinal and Hall conductivities within the Kubo formulas at zero temperature with the vertex corrections for the velocity operator. Vertex corrections are important for describing transport properties in the strong disorder regime. In the case of strong disorder, the longitudinal conductivity is weakly dependent on the disorder strength, while the Hall conductivity decreases with increasing disorder. 
\end{abstract}
\maketitle
\section{Introduction}


The study of the electronic structure of the magnetic materials with the non-trivial topology in a momentum space attracts great interest in modern condensed matter physics~\cite{Chiu2016}. One of the manifestations of the non-trivial topology is the intrinsic anomalous Hall effect (AHE), which arises due to the finite value of the Berry curvature~\cite{Nagaosa2010}. In the insulating state at zero temperature, the Hall conductivity is quantized, leading to the quantum anomalous Hall effect~\cite{Liu2016} (QAHE).

In topological insulators, the non-trivial topology of the bulk band structure leads to the formation of the robust surface states with the Dirac cone~\cite{Hasan2010}. At the charge neutrality point finite out-of-plane magnetization opens a gap in the spectrum leading to the QAHE ~\cite{Wang2015, Tokura2019}. If the chemical potential is outside the gap, then the system is in the AHE regime, which has been extensively studied in the literature~\cite{Nunner2007,Sinitsyn2007,Chiba2017,Sabzalipour2019}.
Experimentally, the chemical doping of topological insulators of the Bi$_2$Se$_3$ family with transition metal elements (Fe, Cr or Mn) induces a bulk magnetisation that opens a mass gap at the Dirac point of the surface states~\cite{Chang2013,Checkelsky2014,Chang2015}. In this case, QAHE is realised with the perfectly quantized Hall conductivity $\sigma_{xy}=\sigma_0=e^2/h$. In real samples, non-uniform doping leads to spatial variations of the mass gap and the chemical potential. Spectroscopy experiments show the important role of scalar and magnetic disorder in the electronic properties of magnetic topological insulators~\cite{Lee2015}. In Refs.~\cite{Liu2020,Satake2020} it was shown that increasing the disorder leads to the transition from the QAHE to the AHE regime. In the disordered samples the Hall conductivity is no longer quantized and finite longitudinal conductivity is observed even at the charge neutrality point. This is due to the presence of the finite density of states which has been observed by spectroscopic~\cite{Lee2015} and angle resolved photon emission spectroscopy measurements~\cite{Lu2021}. The interest in studying the effect of disorder on the magnetic properties of topological insulators arises from possible spintronic applications~\cite{Tokura2019}. The effects of disorder are also crucial for the problem of realising the edge Majorana fermions in the topological insulator-superconductor heterostructure~\cite{Ji2018, Huang2018}.  

In magnetic topological insulators the effects of disorder are usually considered in a weak scattering approximation where the real part of the self-energy is neglected~\cite{Crepieux2001,Nunner2007,Chiba2017}. In Refs.~\cite{Okugawa2020,Sbierski2021} the influence of strong disorder on the electronic properties of the surface state of a topological insulator with magnetization was studied. It was shown that sufficiently strong disorder leads to the finite density of states at the charge neutrality point. Note that for the anomalous Hall effect diagrams beyond the Born approximation series also give a significant contribution~\cite{Ado2016}.

In Ref.~\cite{Klier2019}, the effects of strong disorder on the conductivity of the Weyl semimetal were considered within the self-consistent Born approximation (SCBA). It was shown that SCBA can analytically capture the qualitative behaviour of various observables and their universal features for any disorder strength. 

In this work we study the longitudinal and Hall conductivity of the surface states of the topological insulator with out-of-plane magnetization within the Bastin-Kubo-Streda formulas with vertex corrections. The influence of the random point-like magnetic and scalar perturbations is studied within the high-order Born approximation. We compare the direct summation of the Born series for the self-energy with the analytical results of the SCBA solution for arbitrary values of the disorder strength. We find that disorder reduces the magnetization induced gap and drives the system towards the AHE regime. For small disorder, the density of states vanishes at the Dirac point and the Born series for the self-energy converges monotonically to one branch of the SCBA solution. For sufficiently large disorder $j>j_c$ a finite density of states is generated at the charge neutrality point even for the finite gap in the spectrum. In this case the Born series converges non-monotonically to the other branch of the SCBA solution. For sufficiently large disorder, the Born series for the self-energy loses its causality properties or diverges depending on the values of the parameters. We found that vertex corrections for both retarded-advanced and retarded-retarded vertices are important for describing the longitudinal conductivity at large values of disorder. The Hall conductivity also depends significantly on the vertex corrections. For large disorder, the Hall conductivity decreases with increasing disorder. We compare our results with experimental data.

\section{Hamiltonian}
We consider Hamiltonian of the surface states of the topological insulator with the out-of-plane magnetization in the form~\cite{Hasan2010}
\begin{equation}
H=v(s_xk_y-s_yk_x) +Bs_z-\mu.
\end{equation}
Here $s_i$ are spin Pauli matrices, $v$ is the Fermi velocity of the surface states, $k_{x,y}$ is the momentum of the surface states, $B$ is the value of the out-of-plane magnetization. We choose $(x,y)$ as in-plane directions while $z$ is out-of-plane direction. Spectrum of this Hamiltonian is given by $\epsilon_{\pm}=-\mu\pm\sqrt{B^2+v^2k_x^2+v^2k_y^2}$. QAHE regime corresponds to $\mu<|B|$ where Hamiltonian is fully gapped and Hall conductivity is quantized. Case of $\mu\geqslant |B|$ corresponds to the AHE state with the finite Fermi surface. 

Bare advanced/retarded Green's function of this Hamiltonian are given by
\begin{equation}\label{bare_green}
G_0^{\pm}=\frac{\mu\pm i0+Bs_z+v(s_xk_y-s_yk_x)}{(\mu\pm i0)^2-B^2+v^2k_x^2+v^2k_y^2}.    
\end{equation}

\section{Self-energy}
We consider two cases of short-range random disorder. One is the density disorder corresponding to the randomly distributed charge impurities, the other is the paramagnetic (which we will call magnetic for convenience) with the out-of-plane $s_z$ magnetization corresponding to the spatial fluctuations of the magnetization.
We will describe the disorder by a local potential $\hat{U}_i=u_i s_i \sum\limits_i \delta(\mathbf{r}-\mathbf{R}_j)$, where $\delta(\mathbf{r})$ is the Dirac delta function, $\mathbf{R}_j$ are the positions of the randomly distributed point-like impurities with the local potential $u_i$, and $s_i$ is the Pauli matrix of the potential corresponding to the type of disorder. We assume that the perturbation is Gaussian and uncorrelated, i.e, $\langle \hat{U}_i \rangle=0$ and $\langle \hat{U}_i(\mathbf{r}_1) \hat{U}_j(\mathbf{r}_2) \rangle=n u_0^2 \delta (\mathbf{r}_1-\mathbf{r}_2)\delta_{jl}$, where $\delta_{jl}$ is the Kronecker delta symbol. Scalar disorder corresponds to random local charge with potential $\hat{U}_0=\hat{1}u_0$, while magnetic disorder describes random local magnetization $\hat{U}_z=s_zu_0$. 

Strength of the disorder is given by the dimensionless value $j=n_iu_0^2/(2\pi v^2)$. Weak disorder corresponds to $j \ll 1$ while case $j \sim 1$ corresponds to the strong disorder regime.

We consider self-energy in a Born approximation. In this case we can calculate m-th order of the Born series as a recursive series
\begin{equation}\label{born_eq}
\begin{cases} 
\hat{\Sigma}^{(m+1)}= \sum\limits_{k,i} \langle \hat{U}_i G(\hat{\Sigma}^{(m)}) \hat{U}_i \rangle, \\
G^{-1}(\hat{\Sigma}^{(m)})=-H-\hat{\Sigma}^{(m)}.\end{cases}
\end{equation}

If this series converges $\hat{\Sigma}^{(m)}\rightarrow \hat{\Sigma}$ for $m \rightarrow +\infty$ then the sum of the Born series can be represented as SCBA solution $\hat{\Sigma}= \sum_{k,i} \langle U_i G(\hat{\Sigma}) U_i \rangle$. 

We found that self-energy for the considered types of the disorder has a non-trivial spin structure $\hat{\Sigma}= \Sigma_0 + \Sigma_z s_z$ due to finite magnetization. Explicit expression of self-energy as a sum of $m$ orders of the Born series is
\begin{eqnarray}\label{born_series}
\Sigma_0^{(m+1)}\!&=&\!\!j\frac{\Sigma_0^{(m)}-\mu}{2}\Xi^{(m)}, \,\,
\Sigma_z^{(m+1)}\!=\!-j\frac{\Sigma_z^{(m)}+B}{2}\Xi^{(m)}, \nonumber\\
\Xi^{(m)} = &\ln&\frac{v^2k_c^2}{(B+\Sigma_z^{(m)})^2-(\Sigma_0^{(m)}-\mu)^2}.
\end{eqnarray}
 Initial condition for this self-energy series $\Sigma_0^{(0)}=-i0$ gives advanced self-energy $\hat{\Sigma}^{+}$ that corresponds to the advanced Green's function $G^{+}$. Also, $\Sigma_0^{(0)}=+i0$ gives retarded self-energy $\hat{\Sigma}^{-}$ that corresponds to the retarded Green's function $G^{-}$. We set $\Sigma_z^{(0)}=0$ in our calculations. We will discuss the consequence of other initial conditions in following subsection~\ref{subsec::logmap}.

 Real part of the self-energy leads to the renormalization of the chemical potential $\mu \rightarrow \bar{\mu}=\mu- \textrm{Re}\, \Sigma_0$ and magnetization $B \rightarrow \bar{B}=B+ \textrm{Re}\, \Sigma_z$. Self-energy acquire finite imaginary part Im$\,\hat{\Sigma}^+=-i\Gamma_0-i\Gamma_z s_z$. Impurity averaged Green's function $G^{\pm}$ can be obtained from $G_0$ and self-energy using second equation in Eq.~\eqref{born_eq}. It leads to the renormalization $\mu \rightarrow \bar{\mu} \pm i\Gamma_0$ and $B \rightarrow \bar{B} \mp i\Gamma_z$ in the Green's function given by Eq.~\eqref{bare_green} that leads to following impurity-averaged Green's function
\begin{equation}\label{green}
G^{\pm}=\frac{\bar{\mu} \pm i\Gamma_0+(\bar{B} \mp i\Gamma_z)s_z+v(s_xk_y-s_yk_x)}{(\bar{\mu} \pm i\Gamma_0)^2-(\bar{B} \mp i\Gamma_z)^2+v^2k_x^2+v^2k_y^2}. 
\end{equation}
 
\subsection{Properties of the finite order Born series}
We start our analysis of the disorder with the first Born approximation $m=1$. In this case the self-energy only has a real part if the chemical potential is within the gap $\mu<B$. Direct calculations show that $\Sigma_0^{(1)}=-j\mu \ln{v^2k_c^2/(B^2-\mu^2)}$ and $\Sigma_z^{(1)}=-jB\ln{v^2k_c^2/(B^2-\mu^2)}$. From the Dyson equation $G^{-1}=-H-\Sigma$ we see that disorder renormalises the chemical potential $\bar{\mu}=\mu-\Sigma_0^{(1)}$ and the magnetisation $\bar{B}=B+\Sigma_z^{(1)}$. This renormalisation tends to increase the chemical potential $\bar{\mu}$ and decrease the value of the magnetisation $\bar{B}$. If the disorder is large enough, $\bar{\mu}>\bar{B}$, the gap is closed and a disorder-induced transition from the insulating to the metallic state occurs. 

\begin{figure}[b]
\includegraphics[width=1.0\linewidth]{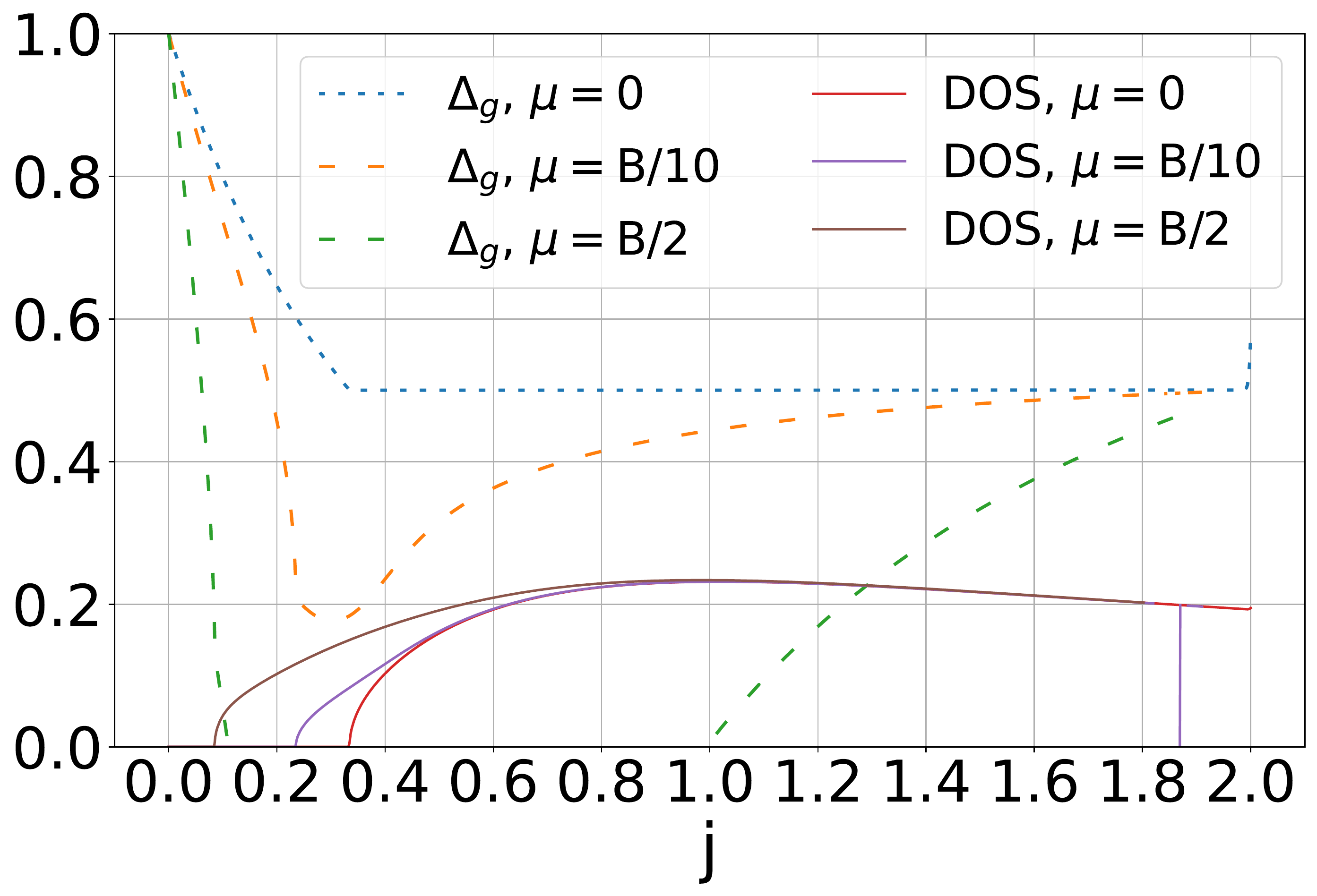}
\caption{Normalized value of the gap $\Delta_g=(\bar{B}-\bar{\mu})/(B-\mu)$ versus disorder strength $j$ is shown as dashed line for different values of the chemical potential $\mu=0$, $\mu=B/10$, $\mu=B/2$. Solid lines represent normalized DOS $=(v/k_c\pi)\textrm{Im Tr} G$ for corresponding values of the chemical potential. Sudden changes in plots for $j \sim 2$ appears due to divergence of the self-energy. We set $vk_c/B=10$. }
\label{critical_disorder}
\end{figure}

We now extend our calculations to the large order of the Born series. We plot the normalised value of the gap $\Delta_g=(\bar{B}-\bar{\mu})/(B-\mu)$ as a function of the disorder j in Fig.~\ref{critical_disorder} for $m=4000$. In this figure we also show the normalised density of states DOS $=(v/k_c\pi)\textrm{Im Tr} G\propto \Gamma_0$. We see that increasing the disorder decreases the gap. Even for small deviations of the chemical potential from the Dirac point, increasing the disorder leads to a rapid decrease in the gap. At $j>j_c= 1/\ln (2vk_c/B)$ a finite density of states appears for $\mu=0$ while the gap remains open. Increasing the chemical potential decreases the value of the disorder where the finite density of states appears inside the gap. For $\mu>B/2$ the density of states appears and the gap closes at similar disorder strengths. We note that for $\mu>B/2$ the value of the gap is quite sensitive to the disorder strength.

Important question for the validity of our results in the convergence of a procedure given by Eq.~\eqref{born_series}. We plot the self-energy as a function of the number of terms in the Born series $m$ for different values of the disorder $j$ for $\mu=B/4$. We find three regimes of convergence of these series. In the case of $j<j_c$ the Born series converges monotonically with the increase of $m$ starting from the first Born approximation $m=1$. In the case of $j>j_c$ we have a non-monotonic behaviour of the self-energy for small values of $m$. Convergence of the self-energy appears for sufficiently large $m$. Increasing the disorder $j$ increases the value of $m$ where the self-energy starts to converge. If $j>j_d$ then the Born series diverges to infinity.

In principle the values of $j_c$ and $j_d$ depend on the values of the chemical potential $\mu$ and the magnetisation $B$. In the case of $\mu=0$ we get the following analytical values of $j_c=1/\ln (2vk_c/B)$ and $j_d=2$.
\begin{figure}[b]
\includegraphics[width=1\linewidth]{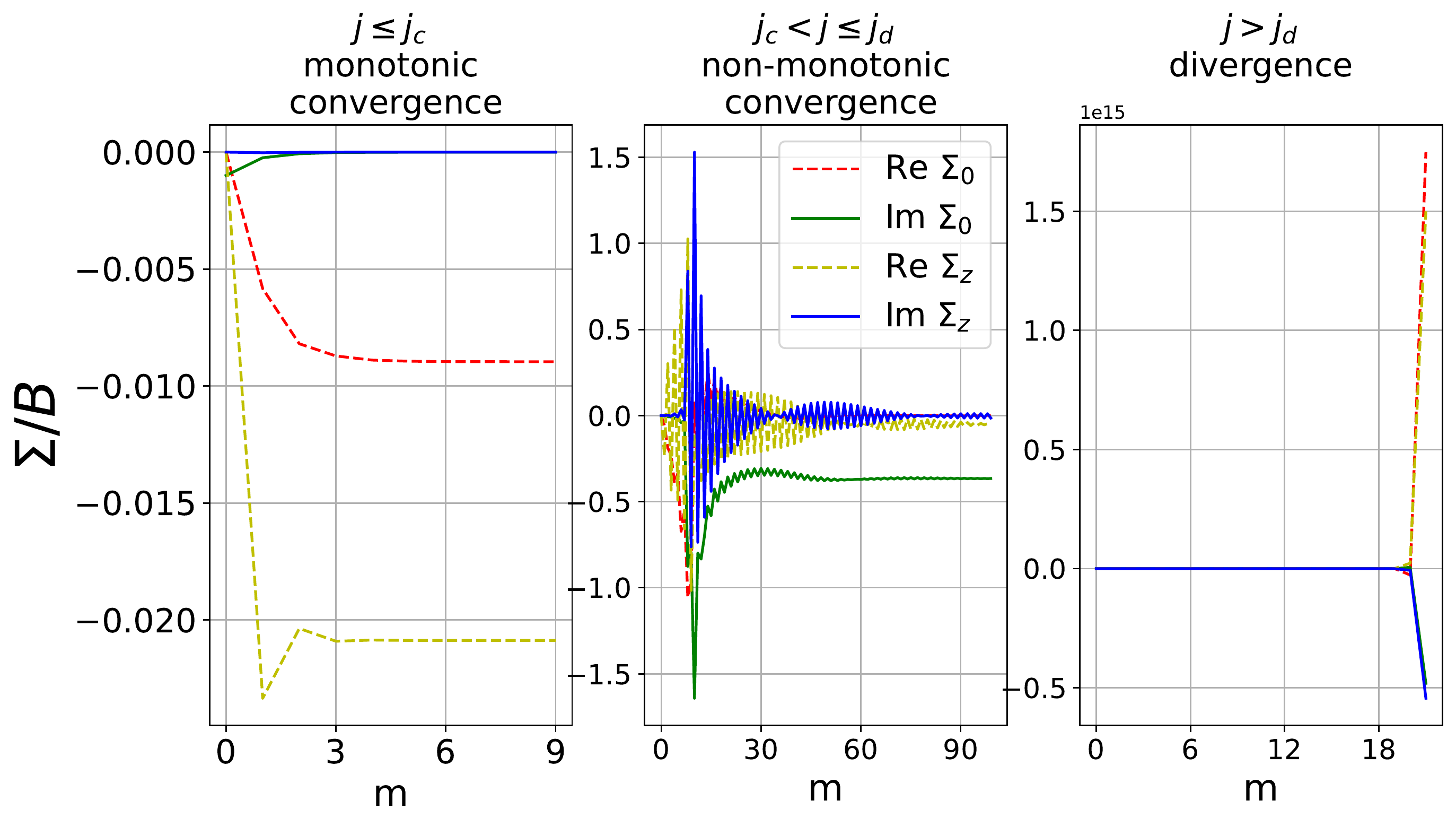}

\caption{Self-energy components $\Sigma/B$ as a functions of a Born series order $m$ for $vk_c/B=10$, $\mu=B/4$. Left figure corresponds to the case of monotonic convergence $j=0.1$. Central figure corresponds to the case of non-monotonic convergence $j=1$. Right figure show divergence of the Born series for $j=2.1$. }
\label{sigma_m0}

\end{figure}

\subsection{Comparison with the analytical SCBA solution}
Now we compare numerical results with the analytical solution of SCBA at Dirac point $\mu=0$. SCBA solution can be obtained from Eq.~\eqref{born_series} if we assume convergence of the self-energy $\Sigma^{(m)} \rightarrow \Sigma$ for $m \rightarrow +\infty$. In this case we have system of equations that determine SCBA self-energy
\begin{equation}\label{scba_analytics}
\begin{cases} 
\Sigma_0=\Sigma_0 \frac{j}{2}\ln \frac{v^2k_c^2}{(B+\Sigma_z)^2-\Sigma_0^2 },\\
\Sigma_z=-(\Sigma_z +B )\frac{j}{2}\ln \frac{v^2k_c^2}{(B+\Sigma_z)^2-\Sigma_0^2 }.\end{cases}
\end{equation}

The first equation in Eq.~\eqref{scba_analytics} has two solutions. The first solution is trivial $\Sigma_0^I=0$. If we insert this into the second equation we have that self-energy has only the real part. Since logarithm is the slow function we can assume first that expression inside the logarithm does not depend on the self-energy and get $\Sigma_z$.  In this case we get first SCBA solution

\begin{equation}\label{first_case_scba}
\Sigma_z^{I}=-\frac{jB\ln \frac{vk_c}B}{1+j\ln \frac{vk_c}B},\quad 
\Sigma_0^I=0.
\end{equation}
We can recalculate self-energy again using the renormalization $B\rightarrow B+\Sigma_z^{I}$ in the logarithm of Eq.~\eqref{first_case_scba} to get more accurate result.

We can see, that self-energy of the first SCBA solution renormalizes real part of the magnetization and imaginary part of the self-energy is infinitesimally small.

Second solution of the first equation in Eq.~\eqref{scba_analytics} is given by $1=\frac{j}{2}\ln \frac{v^2k_c^2}{(B+\Sigma_z)^2-\Sigma_0^2 }$. We insert this expression into the second equation and get $\Sigma_z=-(B+\Sigma_z )$ that gives us $\Sigma_z=-B/2$. If we put it back to the first equation we get that $\Sigma_0= \pm i \sqrt{v^2k_c^2 e^{-\frac{2}{j}}-B^2/4}$. Only one solution conserves causality that leads to the second SCBA solution
\begin{eqnarray}\label{second_case_scba}
\Sigma_z^{II}=-\frac{B}2,\,\,
\Sigma_0^{II}=-i \sqrt{v^2k_c^2 e^{-\frac{2}{j}}-\frac{B^2}4}
\end{eqnarray}
This solution renormalizes real part of the magnetization and generates imaginary part of the self-energy for $j>j_c=1/\ln (2vk_c/B)$. Note, that imaginary part of the self-energy growth exponentially with the increase of the disorder strength. 

\begin{figure}[b]
\includegraphics[width=1\linewidth]{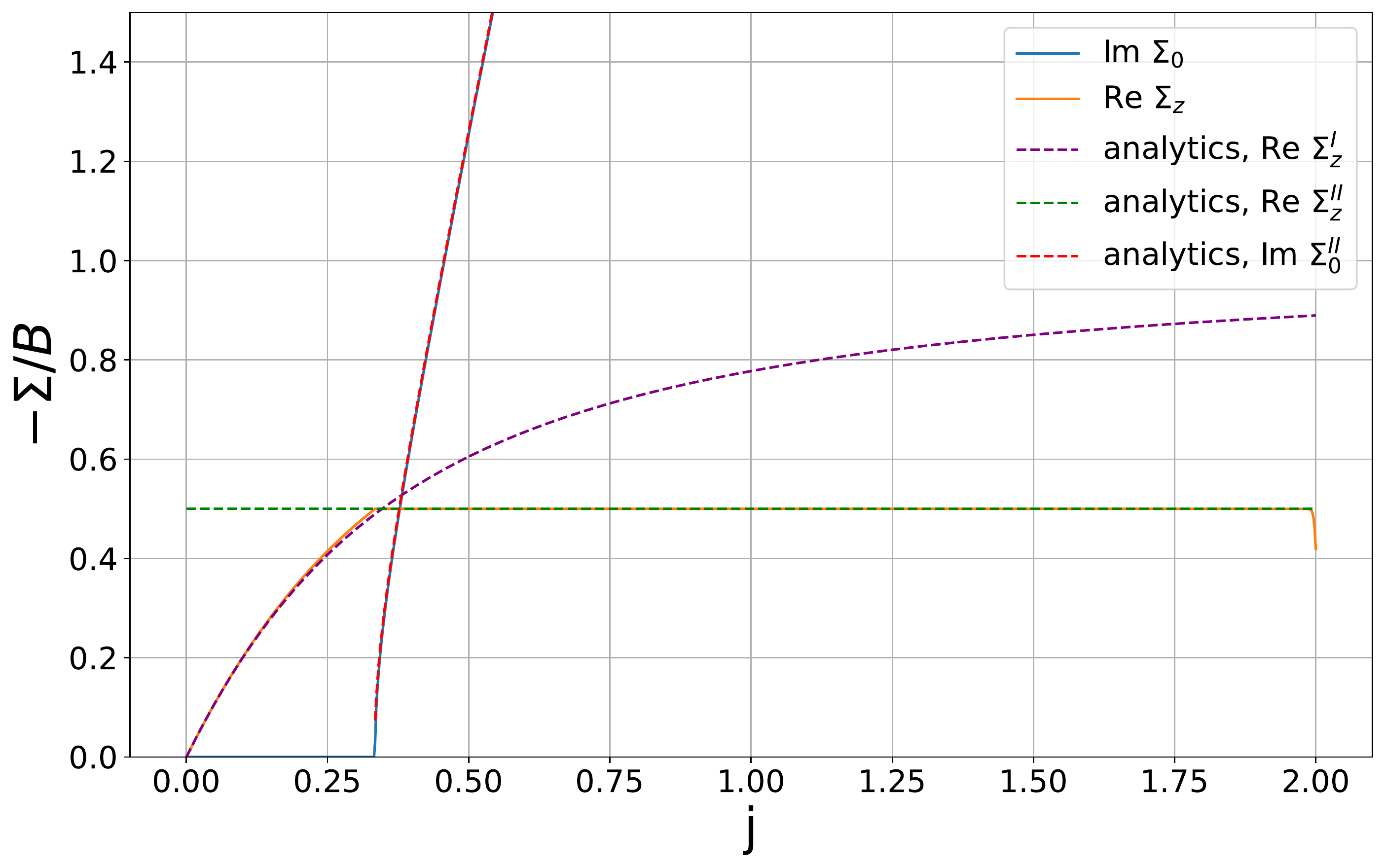}

\caption{Comparison of the analytical SCBA results (dashed lines) for the self-energy with the numerical results (solid lines) for $vk_c/B=10$, $\mu=0$.}
\label{sigma_a}

\end{figure}

We plot Born series self-energy at $\mu=0$ for $m=4000$ and compare it with the SCBA solutions at Fig.~\ref{sigma_a}. We see that for $j<j_c$ self-energy is given by Eq.~\eqref{first_case_scba}. For $j>j_c$ solution for the self-energy is given by Eq.~\eqref{second_case_scba}. Note, that for $j>j_d=2$ Born series diverges, while SCBA solution is always finite regardless the value of $j$.

We can see that with the increase of the disorder imaginary part of the self-energy grows very fast. Even at moderate values of the disorder $j \sim 0.5$ imaginary part of the self-energy exceeds value of the magnetization. It means that for strong disorder imaginary part of the self-energy becomes a dominant energy parameter $\Gamma_0 \gg \mu,B$.

\subsection{Connection between with the logistic map. Lyapunov exponents}\label{subsec::logmap}
\begin{figure}[b]
\includegraphics[width=1\linewidth]{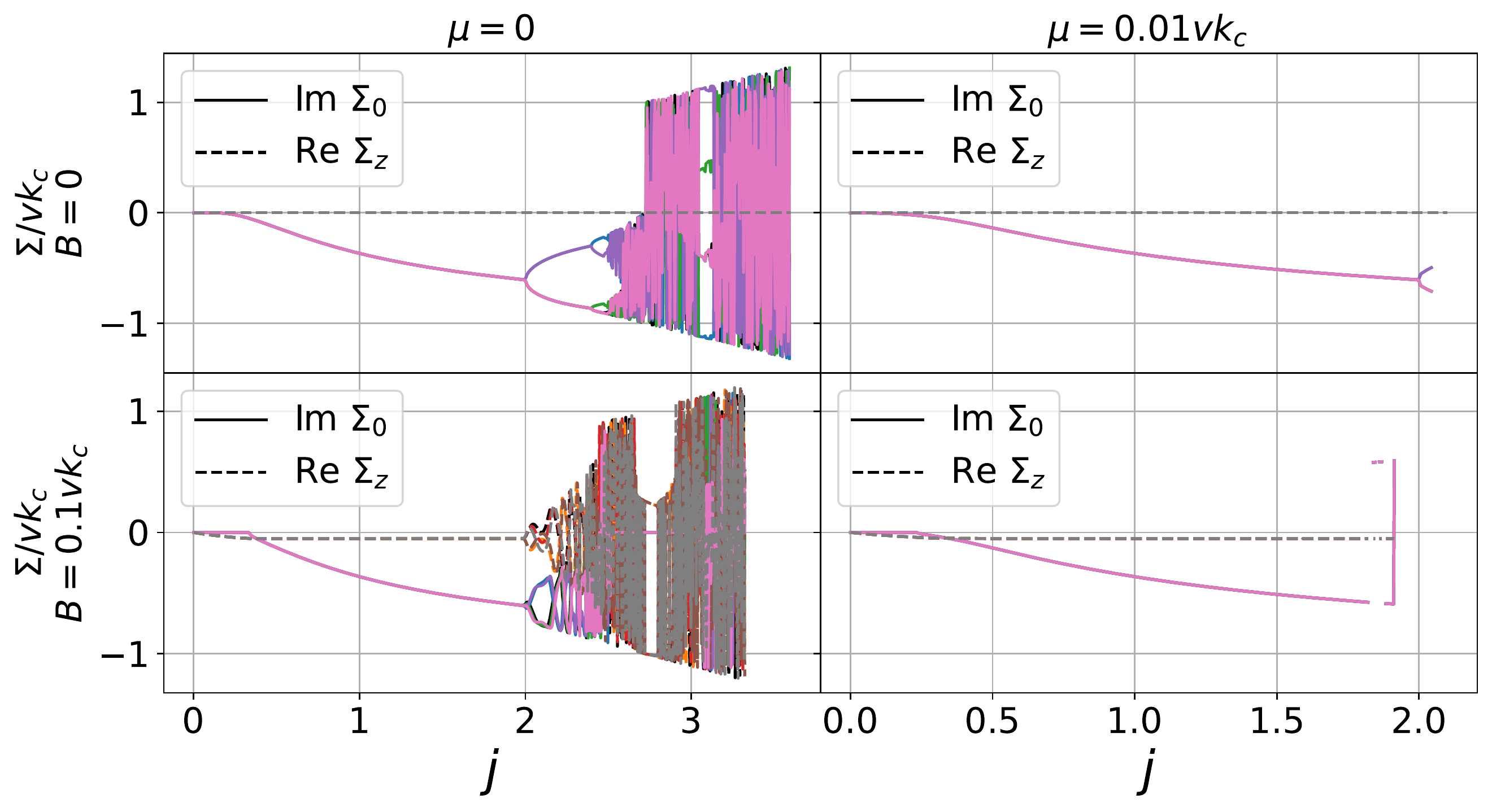}
\caption{Self-energy components $\Sigma_0^{(m)}/vk_c$ and $\Sigma_z^{(m)}/vk_c$ for $m=1000,1001,...,1005$ as a function of the disorder strength $j$ for different values of magnetization $B$ and chemical potential $\mu$. Left column corresponds to $\mu=0$, right column to $\mu=0.01vk_c$. Upper row corresponds to $B=0$, lower row to $B=0.1vk_c$.}
\label{sigma_j}

\end{figure}

We plot numerical solutions for the self-energy $\Sigma^{(m)}$ for different values of $m$ in Fig.~\ref{sigma_j}. We can see that there is no convergence for $j>2$. First we look at the case $\mu=0,B=0$. For $2<j<2.4$ the self-energy oscillates between two branches. Further increases in $j$ increase the number of branches and at some point the behaviour becomes chaotic. Further increase of the disorder strength leads to the loss of the causality of the Green's function (imaginary part of the self-energy changes sign). The case of finite $B$ shows a qualitatively similar behaviour. In the case of $\mu>0$ the solutions quickly diverge to infinity for $j>j_d$ as the Born series order $m$ increases. Such observations suggest chaotic behaviour for large values of the disorder strength. The value of the critical disorder $j_d$ where Born series start to diverge decreases from its maximum value $j_d(\mu=0)=2$ with increasing chemical potential $\mu$.

In order to get insights about the behaviour of the Born series, we write it down at the Dirac point $\mu=0$ without magnetization $B=0$ as
\begin{equation}
 \Sigma_0^{(m+1)}=j\Sigma_0^{(m)}\frac{1}{2}\ln{\frac{v^2k_c^2}{-(\Sigma_0^{(m)})^2}} .  
\end{equation}
Since the real part of the self-energy vanishes we rewrite this equation as
\begin{equation}\label{logmap}
g^{(m+1)}=-jg^{(m)} \ln g^{(m)}, 
\end{equation}
where we introduce dimensionless self-energy as $\Sigma^{(m)} = -i g^{(m)} v k_c$. We will call this series as logarithmic logistic map. 

If this series converges then we have SCBA solution $g_{\textrm{scba}}= e^{-1/j}$ for $m\rightarrow +\infty$. We see that for $j \rightarrow +\infty$ SCBA self-energy approaches energy cut-off $g^{(m)} \rightarrow 1$. If we expand Eq.~\eqref{logmap} by powers of $g^{(m)} - 1 $ get in the lowest order 
\begin{equation}\label{logisticmap}
g^{(m+1)}=jg^{(m)} (1-g^{(m)}).
\end{equation}
This equation is called as logistic map that describes e.g. population growth and is used as a prototypical model for a chaos~\cite{May1976}. Logistic map show a bifurcation behavior for large $j$. It means that logarithmic logistic map given by Eq.~\eqref{logmap} has bifurcation behaviour as well for large $j$.

One way to quantify chaotic behaviour is to calculate local Lyapunov exponents~\cite{Abarbanel1992}. In case of the one dimensional series $x^{(m+1)}=f(x^{(m)})$ Lyapunov exponent is $\lambda=\lim_{m \rightarrow \infty} 1/m \sum_{0}^m \ln |L_m|$ where $L_m = f'(x^{(m)})$. For the system of equations we have spectrum of Lyapunov exponents 
 with $\lambda_i=\lim_{m \rightarrow \infty} 1/m \sum_{0}^m \ln |L_{i,m}|$ where $L_{i,m}$ is the i-th eigenvalue of the matrix $J_{ij}=\partial f_i/\partial x_j$. Lyapunov exponents show how sensitive is the solution to the initial conditions. If Lyapunov exponents are all negative then small perturbations of the initial conditions do not affect solution for $m \rightarrow \infty$. Zero maximal Lyapunov exponent show presence of the degeneracy of solutions: small perturbations in the lead to different stable solutions. If maximal Lyapunov exponent is positive then we have chaotic behaviour and the solution for $m \rightarrow \infty$ is unstable.

\begin{figure}[b]
\includegraphics[width=1\linewidth]{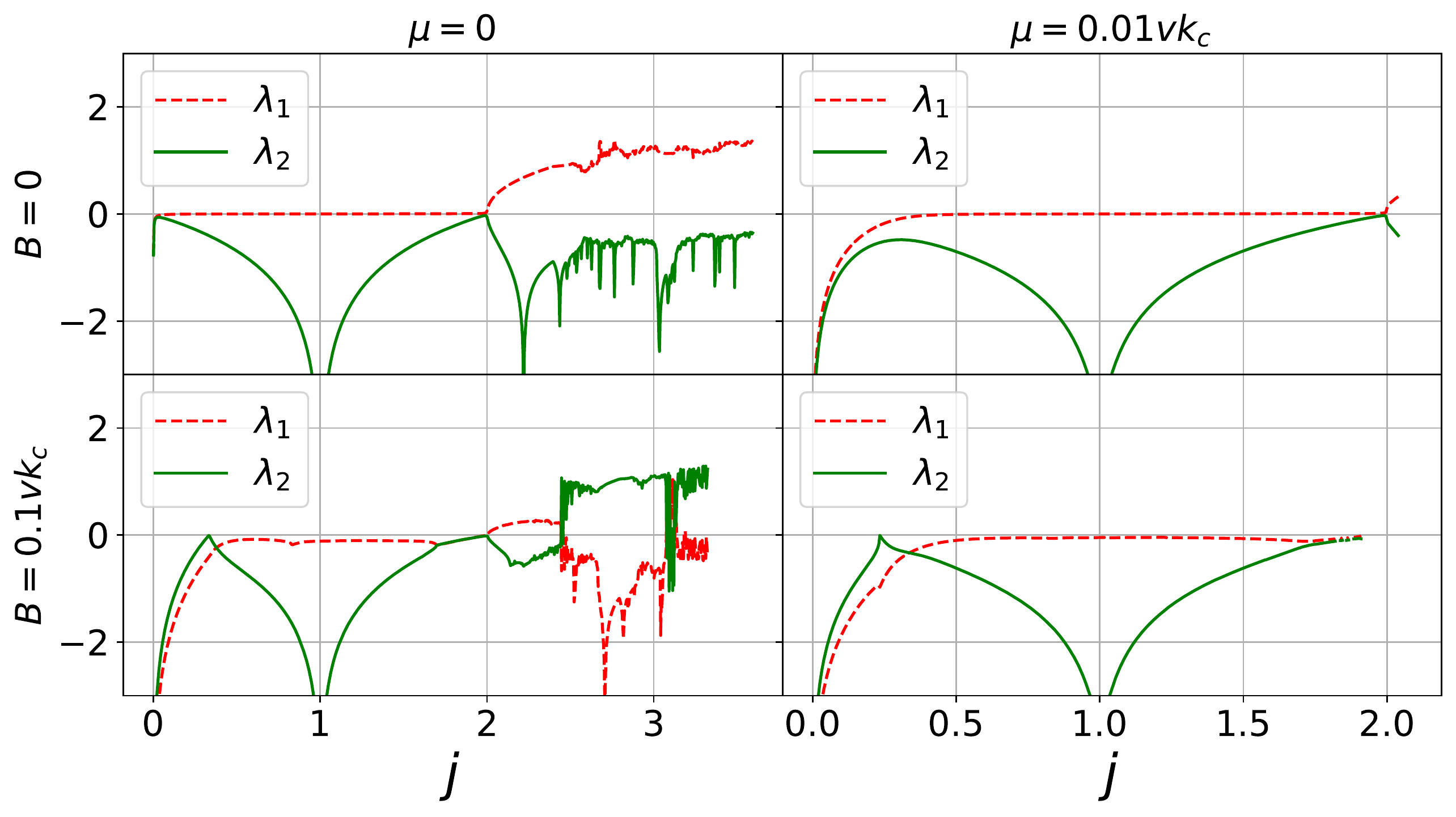}

\caption{Local Lyapunov exponents for the Eq.\ref{born_series} as functions of the disorder strength $j$ for different values of the chemical potential $\mu$ and magnetization $B$. We set $m=1000$. Left column corresponds to $\mu=0$, right column to $\mu=0.01vk_c$. Upper row corresponds to $B=0$, lower row to $B=0.1vk_c$.}
\label{lyap}

\end{figure}

The local Lyapunov exponents for Eq.~\eqref{born_series} are shown in Fig.~\ref{lyap}. We can see that the maximum Lyapunov exponent becomes positive when the Born series starts to diverge. In the case of $B>0$ at $j \sim 0.26$ the maximum Lyapunov exponent approaches zero value, which indicates the transition from the solution given by Eq.~\eqref{first_case_scba} to another SCBA solution given by Eq.~\eqref{second_case_scba}. 

One of the Lyapunov exponents has a zero plato $\lambda_1 \simeq 0$. This is due to the sensitivity of the solution to the initial conditions. Consider the case $\mu=0, B=0$. If the imaginary part $\Gamma_0^{(0)}>\Gamma_z^{(0)}$ we have the solution $\Gamma_0 >0, \Gamma_z=0$ for $m \rightarrow \infty$ for the eq.\ref{born_series}. However, in the case of $\Gamma_0^{(0)}<\Gamma_z^{(0)}$ we have another solution $\Gamma_0 =0, \Gamma_z>0$. So we have two stable solutions for two different types of initial conditions, either $\hat{\Sigma}^{(0)} = -i0$ or $\hat{\Sigma}^{(0)} = -i0s_z$, where $i0$ is the numerically small value. Usually in condensed matter physics the initial condition for the self-energy is taken to be $\hat{\Sigma}^{(0)} = -i0$. Note that for eq.~\eqref{logisticmap} we will not have such a zero Lyapunov exponent, since we have already assumed $\Gamma_z^{(0)}=0$ in this equation, and consequently we have only one Lyapunov exponent.

\section{Vertex corrections}
The introduction of self-energy into the current-current correlation function violates gauge invariance. To restore the gauge invariance we have to introduce the vertex corrections to the velocity operator~\cite{Chiba2017}.

We can find corrections to the velocity operator in the form $\delta v_{I(II)\alpha}=V_{I(II)\alpha}-v_{\alpha}$. Thus we can find the $n$-th order of vertex corrections as $\delta v_{I(II)\alpha}^{(n)}= \sum\limits_{k,i}\langle \hat{U}_i G^+ (v_{\alpha}+\delta v_{I(II)\alpha}^{(n-1)})G^{-(+)} \hat{U}_i \rangle$. Here $I$ corresponds to the retarded-advanced vertex correction $\langle G^+ ... G^-\rangle$, while $II$ corresponds to the advanced-advanced vertex correction $\langle G^+ ... G^+\rangle$. We found that in our model the vertex corrections always converge when the Born series of the self-energy converges. There is also always a single solution for the SCBA vertex corrections.

Since we consider point-like disorder, the vertex corrections have no $k$ dependence. Calculation of the vertex corrections for retarded-advanced Green's functions in a first Born approximation $V^{(1)}_{Ix}=v_x+ n_iu_0^2 \sum_k G_0^+ v_x G_0^-$ shows that vertex corrections renormalise the value of the bare velocity operator and introduce a new $s_x$ term. We introduce a renormalisation of the current operator in the form $\delta v_{Ix}=V_{Ix}-v_x$. In this case vertex corrections can be found as~\cite{Chiba2017} 
\begin{eqnarray}
\delta v_{Ix}=\frac{n_iu_0^2}{(2\pi)^2}\int d^2k G^+ (v_x + \delta v_{Ix}) G^-,
\end{eqnarray}
where Green's function are given by Eq.~\eqref{green}. After we take Tr$[s_i...]$ for the left and right parts of the second equation we get following system of equations for the vertex corrections
\begin{eqnarray}
\delta v_{Ix}=v\delta_xs_x+v\delta_ys_y, \nonumber\\
\delta_x=j(-M_{xy}+\delta_xM_{xx}+\delta_yM_{xy}),\nonumber\\
\delta_y=j(-M_{yy}+\delta_xM_{yx}+\delta_yM_{yy}).
\end{eqnarray}
Here $M_{\alpha\beta}=v^2/(4\pi)\int d^2k \textrm{Tr}[s_{\alpha}G^+s_{\beta}G^-]$. Since this integral converges we use infinite limits in the integration $k_c \rightarrow +\infty$ to get analytical results. Explicit integration gives us
\begin{eqnarray}
M_{xx}&=&\frac{\bar{\mu}^2-\bar{B}^2+\Gamma_0^2-\Gamma_z^2}{4(\bar{B}\Gamma_z+\bar{\mu}\Gamma_0)}(\pi/2-\theta),\nonumber\\
M_{yx}&=&\frac{\bar{B}\Gamma_0+\bar{\mu}\Gamma_z}{2(\bar{B}\Gamma_z+\bar{\mu}\Gamma_0)}(\pi/2-\theta), \nonumber\\
\theta&=&\arctan \frac{\bar{B}^2+\Gamma_0^2-\bar{\mu}^2-\Gamma_z^2}{2(\bar{B}\Gamma_z+\bar{\mu}\Gamma_0)}.
\end{eqnarray}
Using that $M_{xx}=M_{yy}$ and $M_{yx}=-M_{xy}$ we the vertex corrections in the form
\begin{eqnarray}\label{vertexI}
\delta_x=\frac{jM_{yx}}{(jM_{xx}-1)^2+j^2M_{yx}^2}, \\
\delta_y=1+\frac{jM_{xx}-1}{(jM_{xx}-1)^2+j^2M_{yx}^2}. 
\end{eqnarray}

In case of vertex corrections for the retarded-retarded (or advanced-advanced) Green's function we calculate first Born approximation $V_{II1x}=v_x+ n_iu_0^2 \sum_k G^+ v_x G^+$. We found out that vertex corrections renormilize value of the bare vertex $v_x$. In this case we seek renormalization of the current operator in the form $\delta v_{IIx}=V_{IIx}-v_x$
\begin{equation}
\delta v_{IIx}=\frac{n_iu_0^2}{(2\pi)^2}\int d^2k G^+ (v_x + \delta v_{IIx}) G^+  
\end{equation}
We use that $N_{yy}=v^2/(4\pi)\int d^2k \textrm{Tr}[G^+ s_y G^+s_y]=-1/2$ and get that 
\begin{equation}\label{vertexII}
\delta v_{IIy}=v\frac{j}{j+2} s_y
\end{equation}
These results are can be easily generalized for other types of the disorder. In case of the magnetic disorder we get that vertex corrections are given by the same equations up to $j\rightarrow -j$ in Eqs.\eqref{vertexI} and \eqref{vertexII}. 

We start with the case of weak disorder $j<j_c$ for the gapped phase $\Gamma_0\rightarrow 0$. In this case $M_{xx}=-1/2$ and $M_{yx}=-\bar{B}\Gamma_0/(\bar{\mu}^2-\bar{B}^2)\rightarrow 0$. Thus, we get that vertex correction $\delta_y=j/(j+2)$ is finite inside the gapped region. If we compare this result with Eq.~\eqref{vertexII} we get that vertex corrections for $I$ and $II$ contributions are equal $\delta v_{Ix}=\delta v_{IIx}$ in QAHE regime. We show later that this results in the vanishing of the total longitudinal conductivity.

Now we consider case of strong disorder disorder $j>j_c$ with $\Gamma_0 \gg \mu, B$ and $\Gamma_z =0$. That results in $M_{xx}=1/2$ and $M_{yx}=\bar{B}/\Gamma_0$, $\delta_y=-j/(2-j)$ and $\delta_x=4j\bar{B}/(\Gamma_0(2-j)^2)$. We see that vertex corrections change its sign for strong disorder. Vertex corrections diverge for $j>j_d=2$ at the same value of the disorder as self-energy diverges.  
\section{Longitudinal conductivity}
At zero temperature longitudinal conductivity has two contributions~\cite{Crepieux2001}
\begin{eqnarray}\label{spin_surface}
\sigma_{{\alpha}\alpha}^{I}=\frac {e^2}{2\pi}\int \frac{d^2k}{(2\pi)^2} \textrm{Tr} [V_{I\alpha}\, G^+
\,v_{\alpha}  \, G^-] ,\\
\sigma_{{\alpha}\alpha}^{II }=-\frac {e^2}{4\pi}\int \frac{d^2k}{(2\pi)^2}\textrm{Tr} [V_{II\alpha}\, G^+ \,v_{\alpha}  \, G^+ + c.c ].
\end{eqnarray}

Here we consider $\hbar=1$, bare velocity operator is $v_{\alpha}= \partial H /\partial k_{\alpha}$, $V_{I(II)\alpha}$ is the velocity operator with vertex corrections for $I(II)$ conductivity terms, c.c stand for complex conjugation. We decompose conductivity $\sigma_{{\alpha}\alpha}^{I(II)}=\sigma_{{\alpha}\alpha}^{I(II)bv}+\sigma_{{\alpha}\alpha}^{I(II)vc}$ into the contributions from the bare bubble and vertex corrections. We express these terms as
\begin{eqnarray}
\sigma_{xx}^{Ibv}=\frac{\sigma_0}{\pi}M_{yy}, \quad
\sigma_{xx}^{Ivc}=-\frac{\sigma_0}{\pi}(\delta_yM_{yy}-\delta_xM_{yx}), \nonumber\\
\sigma_{xx}^{IIbv}=-\frac{\sigma_0}{\pi}N_{yy}, \quad
\sigma_{xx}^{IIvc}=\frac{\sigma_0}{\pi}\frac{j}{j+2} N_{yy}
\end{eqnarray}
Here $\sigma_0=e^2/(2\pi)=e^2/h$ is the conductivity quanta. Note, that $M_{xx}=M_{yy}$ and $M_{xy}=-M_{yx}.$

We start with the analysis of the gapped state for weak disorder $j<j_c$. In this case $\Gamma_0\rightarrow 0$ and $\Gamma_z=0$. This results in cancellation between $\sigma_{xx}^{I}$ and $\sigma_{xx}^{II}$ terms. In means that total conductivity $\sigma_{xx} = 0$ vanishes exactly in the QAHE regime. 

Now we consider case of strong disorder disorder $j>j_c$ with $\Gamma_0 \gg \mu, B$ and $\Gamma_z =0$. This results in 
\begin{equation}\label{sxx_an}
\sigma_{xx}=\frac{4\sigma_0}{\pi(4-j^2)}.
\end{equation}

\begin{figure}[h]
\includegraphics[width=1\linewidth]{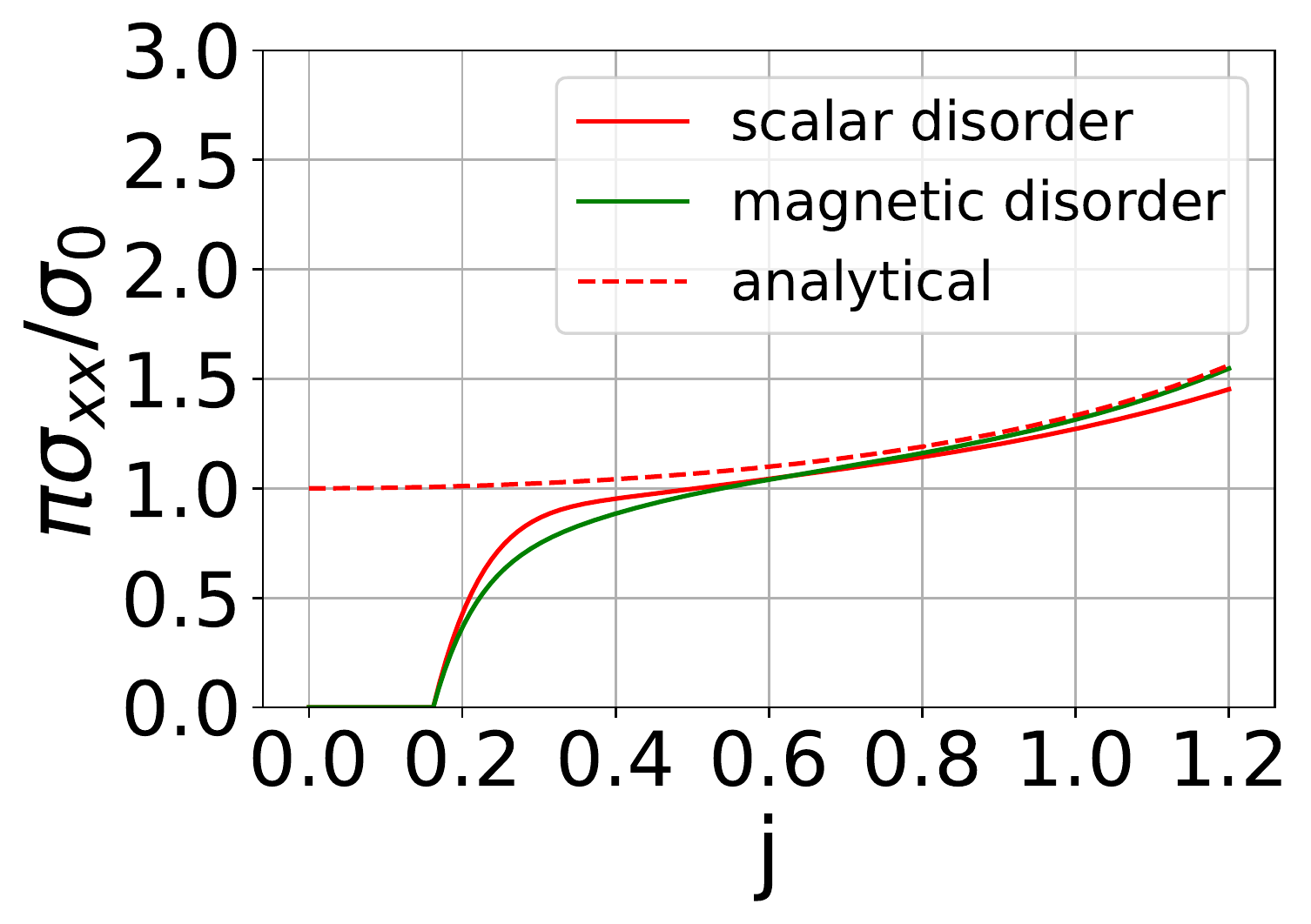}

\caption{Longitudinal conductivity $\sigma_{xx}$ as a function of the disorder strength for $vk_c/B=10$, $\mu=B/4$. Solid lines correspond to the numerical calculations, dashed line is the analytical result given by Eq.~\eqref{sxx_an} }
\label{sxx}

\end{figure}

We see that for a strong disorder longitudinal conductivity has weak dependence on the values of the chemical potential and magnetization. We plot longitudinal conductivity for different values of the disorder strength $j$ at Fig.~\ref{sxx}. In the metallic state value of the longitudinal conductivity approaches value given by Eq.~\eqref{sxx_an} quite quickly. Difference between scalar and magnetic disorders for the value of the conductivity is small.

\section{Hall conductivity}
In general, Hall conductivity has three terms~\cite{Crepieux2001}
	\begin{eqnarray}\label{spin_surface_xy}
\sigma_{{\alpha}\beta}^{I}=\frac {e^2}{2\pi}\int \frac{d^2k}{(2\pi)^2} \textrm{Tr} [V_{I\alpha}\, G^+
\,v_{\beta}  \, G^-] ,\\
\sigma_{{\alpha}\beta}^{II }=-\frac {e^2}{4\pi}\int \frac{d^2k}{(2\pi)^2}\textrm{Tr} [V_{II\alpha}\, G^+ \,v_{\beta}  \, G^+ + c.c ].
\end{eqnarray}
\begin{eqnarray}\label{spin_filled}
\sigma_{{\alpha}\beta}^{III\gamma}=\frac {e^2}{4\pi}\int \frac{d^2k}{(2\pi)^2} \int\limits_{-\infty}^{\mu} f(E) dE \times\\ \nonumber \textrm{Tr} [v_{\alpha}\, G^+\,v_{\beta}  \, \frac {d G^+}{dE}-v_{\alpha}\, \frac {d G^+}{dE}\,v_{\beta}  \,G^+ + c.c.] ,
\end{eqnarray}
Here $\partial G^{\pm}/\partial E=-G^{\pm 2}$ is the derivative of the Green's function over the energy, $f(E)$ is the Fermi distribution function (that is Heaviside step for zero temperature). 

Direct calculation shows that $\sigma_{{x}y}^{II}=0$. Now, we decompose first term in to the contribution from the bare vertex and vertex corrections $\sigma_{xy}^{I}=\sigma_{xy}^{Ibv}+\sigma_{xy}^{Ivc}$. This results in 
\begin{equation}
\sigma_{xy}^{Ibv}=-\frac{\sigma_0}{\pi}M_{yx},\quad
\sigma_{xy}^{Ivc}=\frac{\sigma_0}{\pi}(\delta_yM_{yx}+\delta_xM_{xx}).\\
\end{equation}

As for $\sigma_{xy}^{III}$ there is no compact expression and we will present the analytical results for the total Hall conductivity in the limiting cases. 

In case of $\mu=0$ and $\Gamma_z=0$ expression for the contribution from the filled state becomes especially simple
\begin{eqnarray}
\sigma_{xy}^{III}=\frac{\sigma_0}{2}\left(\frac{2}{\pi}\arctan \frac{\Gamma_0}{\bar{B}}-1\right).
\end{eqnarray}

Now we analyze these expressions for gapped state for weak disorder $j<j_c$ where $\Gamma_0\rightarrow 0$ and $\Gamma_z=0$. In this case we have vanishing contribution from the Fermi surface $\sigma_{xy}^{I}=0$ only contribution comes from the filled states $\sigma_{xy}=\sigma_{xy}^{III}=-\sigma_0/2$. 

\begin{figure}[h]
\includegraphics[width=1\linewidth]{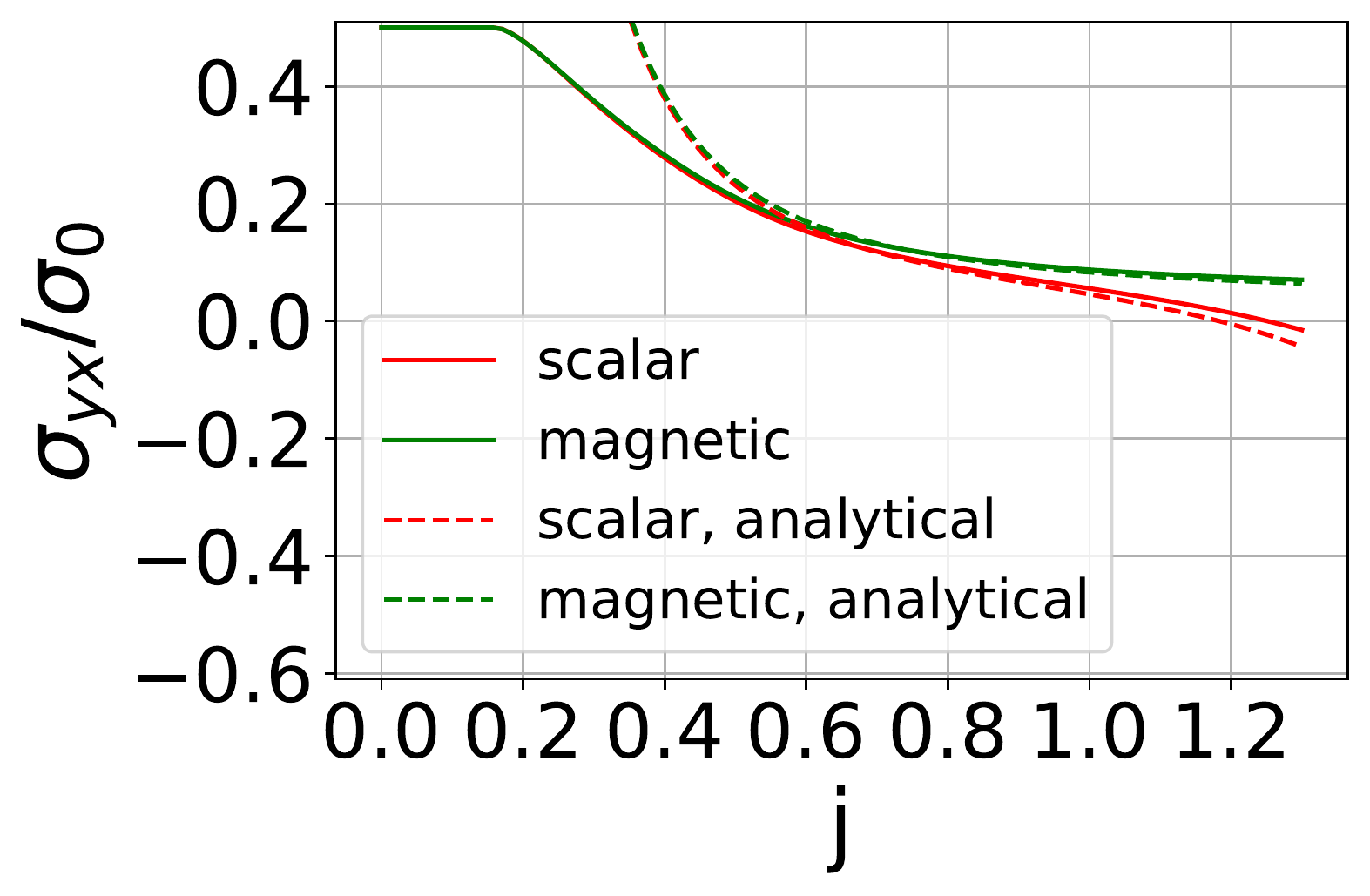}
\caption{Hall conductivity $\sigma_{xy}$ as a function of the disorder strength for $vk_c/B=10$, $\mu=B/4$. Solid lines correspond to the numerical calculations, dashed line is the analytical result given by Eq.~\eqref{sxy_an} }
\label{syx}

\end{figure}

For the strong disorder $j>j_c$ we have $\Gamma_0 \gg \mu, B$ that results in the following expression
\begin{eqnarray}\label{sxy_an}
\sigma_{xy}=-\frac{\sigma_0}{\pi}\frac{\bar{B}}{\Gamma_0} \frac{8-8j+j^2}{(j-2)^2}.
\end{eqnarray}
We plot numerical and analytical results for the Hall conductivity in Fig.~\ref{syx}.
We see that Hall conductivity decreases with the increase of the disorder $j$. At $j \sim 1.17$ Hall conductivity changes its sign due to large contribution from the vertex corrections in case of density disorder. For the magnetic disorder Hall conductivity does not changes its sign.

\section{Conclusions}
In this work we have analysed in detail the properties of the disordered surface states of the topological insulator with out-of-plane magnetization. We find that the linear spectrum of the surface states brings an interesting behaviour of the self-energy arising due to the disorder. Increasing the disorder changes the monotonic convergence of the Born series to non-monotonic. Further increase of the disorder leads to divergence of the Born series.

The Born series is essentially a many-body perturbation series to the single-particle Green's function. The divergence of the Born series shows a failure of the perturbation approach. It means that we should use more advanced techniques beyond perturbation theory to account for disorder effects. The divergence of the perturbation series is not uncommon in condensed matter physics. In the Refs ~\cite{Kozik2015,Gunnarsson2017} it was shown that studying the perturbation series beyond its radius of convergence leads to unphysical results.

We get that the density of states increases exponentially with the increase of the disorder $\rho \propto \exp(-1/j)$ in the metallic state, see Eq.~\eqref{second_case_scba}. Longitudinal conductivity does not follow the Drude formula $\sigma_{xx}\propto \rho$ for strong disorder. Instead, the longitudinal conductivity depends weakly on the disorder strength $\sigma_{xx}\propto \sigma_0/\pi$. Hall conductivity is suppressed but finite. So large density of states $\rho$ together with small longitudinal conductivity $\sigma_{xx}\propto \sigma_0/\pi$ and partially suppressed Hall conductivity $\sigma_{xy} \lesssim \sigma_0/2$ are the hallmarks of the high disorder regime for the magnetised surface state of the topological insulator. In the recent work~\cite{Yano2021} ARPES shows a large density of states at the surface of the ferromagnetic topological insulator. However, transport measurements show a small longitudinal current. Also, the Hall conductivity is suppressed compared to the quantized value. These observations suggest a regime of strong disorder in such samples.

\section*{Acknowledgments}
This work is supported by Russian Science Foundation (project № 22-72-10074). 
\bibliography{disorderedMTI}
\end{document}